\documentclass[twocolumn,showpacs,preprintnumbers,amsmath,amssymb]{revtex4}

\usepackage{graphicx}
\usepackage{dcolumn}
\usepackage{bm}
\usepackage{soul}
\usepackage{color}
\usepackage{epstopdf}
\usepackage[version=3]{mhchem}
\begin{document}

\title{Near-field heat transfer between gold nanoparticle arrays}

\author{ Anh D. Phan$^{1,2}$, The-Long Phan$^{3}$, and Lilia M. Woods$^{1}$}
\affiliation{$^{1}$Department of Physics, University of South Florida, Tampa, Florida 33620, USA}%
\email{anhphan@mail.usf.edu}
\affiliation{$^{2}$Institute of Physics, 10 Daotan, Badinh, Hanoi, Vietnam}%
\affiliation{$^{3}$Department of Physics, Chungbuk National University, Cheongju 361-763, Korea}
\email{ptlong2512@yahoo.com}
\date{\today}

\begin{abstract}
The radiative heat transfer between gold nanoparticle layers is presented using the coupled dipole method. Gold nanoparticles are modelled as effective electric and magnetic dipoles interacting via electromagnetic fluctuations. The effect of higher-order multipoles is implemented in the expression of electric polarizability to calculate the interactions at short distances. Our findings show that the near-field radiation reduces as the radius of the nanoparticles is increased. Also, the magnetic dipole contribution to the heat exchange becomes more important for larger particles. When one layer is displayed in parallel with respect to the other layer, the near-field heat transfer exhibits oscillatory-like features due to the influence of the individual nanostructures. Further details about the effect of the nanoparticles size are also discussed.
\end{abstract}
\maketitle
\section{Introduction}
Noble metallic nanoparticles (MNPs) have been exploited in a wide range of technological applications due to their unique properties. In particular, their strong absorption of radiation together with the ability of control of localized surface plasmon resonances have been key factors in a number of optical devices \cite{1,18}. For many targeted uses and perspectives, periodic two- or three-dimensional MNP arrays have been utilized \cite{1,17,2}. It was shown that many-body effects enhance the electromagnetic behavior of the system compared to the one of the individual particles. As two nanoplasmonic arrays are brought at small separations and maintained at different temperatures, radiative heat transfer occurs. The origin of this exchange process originates from the electromagnetic fluctuations between the objects \cite{6}. Since the electric properties of MNs are sensitive to the external fields \cite{26}, it is possible to employ these fields to change the heat radiation. Much experimental \cite{3,13,14} and theoretical \cite{4,15,16,20,21,22} efforts have been devoted in understanding this phenomenon and finding ways for efficient control.

\section{Theoretical background}
\begin{figure}[htp]
\includegraphics[width=8.5cm]{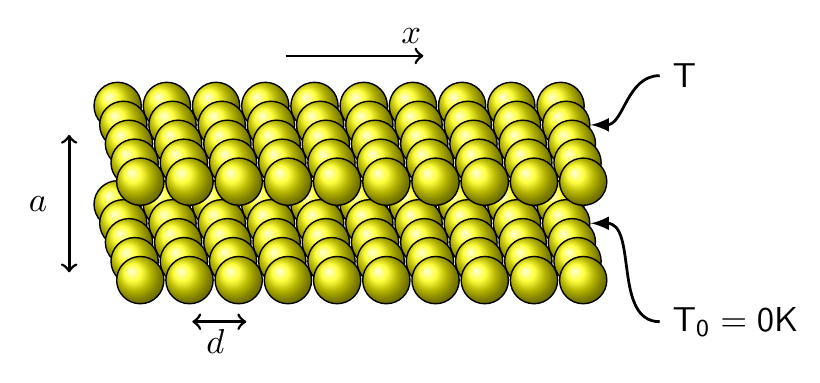}
\caption{\label{fig:0}(Color online) Schematic representation of two layers of gold MNPs kept at temperature $T$ and 0 $K$ on the top and bottom, respectively. The two surfaces are separated by a distance $a$. The separation between centers of adjacent MNPs is $d = 2R + 1$ nm.}
\end{figure}

In this work we focus on the radiative heat transfer between two gold MNP layers. Each nanoparticle is modelled as a dipole. Each layer consists of $20\times 20$ identical particles separated by $1$ nm, as shown in Fig.\ref{fig:0}. It is assumed that each nanoparticle has a spherical shape with radius $R$ and the dielectric and magnetic properties are described via a dipolar model. The radiative heat exchange $P_{i\rightarrow j}(\omega)$ between the $i$-th and $j$-th dipoles consists of electric  $P^e_{i\rightarrow j}(\omega)$ and magnetic $P^m_{i\rightarrow j}(\omega)$ contributions \cite{5,7}, as follows:

\begin{eqnarray}
P^e_{i\rightarrow j}(\omega)=\frac{\omega\varepsilon_0}{\pi}\ce{Im}\alpha^e_j({\omega})\left\langle |\mathbf{E}_{ji}|^2\right\rangle,\nonumber\\
P^m_{i\rightarrow j}(\omega)=\frac{\omega\mu_0}{\pi}\ce{Im}\alpha^m_j({\omega})\left\langle |\mathbf{H}_{ji}|^2\right\rangle,
\label{eq:1}
\end{eqnarray}
where $\alpha^e_j({\omega})$ and $\alpha^m_j({\omega})$ are the electric and magnetic polarizabilities, respectively, of the j-th dipole with an electric $p_j$ and magnetic $m_j$ components. 
Also, $\mathbf{E}_{ji}$ and $\mathbf{H}_{ji}$ are the  electric and magnetic fields, respectively, at position $\mathbf{r}_j$ due to the fluctuations of dipole. $\varepsilon_0$ is the vacuum permittivity and $\mu_0$ is the permeability of free space. The relation between $\mathbf{E}_{ji}$ and the electric dipole moment $\mathbf{p}_{j}$ is given $\mathbf{E}_{ji}(\omega)=\mu_0\omega^2 \mathbf{G}(\mathbf{r}_j,\mathbf{r}_i,\omega)\mathbf{p}_i$ \cite{6,8}. Here  $\mathbf{G}(\mathbf{r}_j,\mathbf{r}_i,\omega)$ is the dyadic Green tensor \cite{8}. Using the fluctuation dissipation theorem \cite{6}, one finds

\begin{eqnarray}
\left\langle \mathbf{E}_{ji}(\omega)\mathbf{E}^*_{ji}(\omega ')\right\rangle &=&\mu_0^2\omega^2\omega '^2\sum_{k,l,t}G_{kl}(\mathbf{r}_j,\mathbf{r}_i,\omega)\nonumber\\
&\times & G^\dagger_{kt}(\mathbf{r}_j,\mathbf{r}_i,\omega ')\left\langle p_{i,l}(\omega)p^*_{i,t}(\omega ')\right\rangle, \nonumber\\
\left\langle p_{i,l}(\omega)p^*_{i,t}(\omega ')\right\rangle &=& \frac{2\varepsilon_0}{\omega}\ce{Im}\alpha^e_i(\omega)\Theta(\omega,T_i)\delta_{lt}\delta(\omega - \omega '),\nonumber\\
\Theta(\omega,T_i) &=& \frac{\hbar\omega}{e^{\hbar\omega/k_BT_i}-1},
\label{eq:2}
\end{eqnarray} 
where $k, l, t = x, y, z$; $\hbar$ is the Planck constant, $k_B$ is the Boltzmann constant, $T_i$ is the temperature of dipole i. 

Solving Eq.(\ref{eq:1}) and Eq.(\ref{eq:2}) together, the exchanged power caused by the electric dipoles is found to be:
\begin{eqnarray}
P^e_{i\rightarrow j}(\omega)&=&\frac{2}{\pi}\frac{\omega^4}{ c^4}\ce{Im}\alpha^e_j(\omega)\ce{Im}\alpha^e_i(\omega)\Theta(\omega,T_i)\nonumber\\
&\times & \ce{Tr}\left(\mathbf{G}(\mathbf{r}_j,\mathbf{r}_i,\omega)\mathbf{G}(\mathbf{r}_j,\mathbf{r}_i,\omega)^\dagger \right),
\label{eq:3}
\end{eqnarray} 
where c is the speed of light. 

Similar considerations apply for the magnetic dipole moments and the magnetic fields, yielding $\mathbf{H}_{ji}(\omega)=(\omega/c)^2 \mathbf{G}(\mathbf{r}_j,\mathbf{r}_i,\omega)\mathbf{m}_i$ \cite{9}.Consequently, the correlation functions for the magnetic dipoles is expressed as \cite{5}
\begin{eqnarray}
\left\langle m_{i,l}(\omega)m^*_{i,t}(\omega ')\right\rangle &=& \frac{2\delta_{lt}}{\omega\mu_0}\ce{Im}\alpha^m_i(\omega)\Theta(\omega,T_i)\delta(\omega - \omega ').\nonumber\\
\label{eq:4}
\end{eqnarray}

Thus the exchanged power due to the magnetic field fluctuations becomes 
\begin{eqnarray}
P^m_{i\rightarrow j}(\omega)&=&\frac{2}{\pi}\frac{\omega^4}{ c^4}\ce{Im}\alpha^m_j(\omega)\ce{Im}\alpha^m_i(\omega)\Theta(\omega,T_i)\nonumber\\
&\times & \ce{Tr}\left(\mathbf{G}(\mathbf{r}_j,\mathbf{r}_i,\omega)\mathbf{G}(\mathbf{r}_j,\mathbf{r}_i,\omega)^\dagger \right).
\label{eq:5}
\end{eqnarray} 

Since particles are taken to be identical, one has $\alpha^{e,m}_1=\alpha^{e,m}_2=...=\alpha^{e,m}_N=\alpha^{e,m}$. It is important to note that since the separation distance between two adjacent gold NPs is not much larger than their radius, the influence of higher-order multipoles (quadrupole in our calculation) on the polarizability of MNPs should be taken into account. We can introduce the effective electric and magnetic polarizabilities for MNPs (R less than the skin-depth) derived from the Mie scattering theory \cite{10,19}
\begin{eqnarray}
\alpha^e(\omega) &=&4\pi R^3\left[\frac{\varepsilon-1}{\varepsilon+2} +\frac{1}{12}\left(\frac{\omega R}{c}\right)^2\frac{\varepsilon - 1}{\varepsilon + 3/2}\right], \nonumber\\
\alpha^m(\omega) &=& \frac{2\pi}{15}R^3\left(\frac{\omega R}{c}\right)^2(\varepsilon-1),
\label{eq:7}
\end{eqnarray}
where $\varepsilon(\omega)$ is the dielectric function of gold NPs. The first and second term in the expression of $\alpha^e(\omega)$ correspond to the dipole and quadrupole contributions, respectively. Authors in Ref.\cite{25} used the dipole term and indicated that the distance between centers of MNPs should be at least few times greater than their radius $R$ to ensure the validity of the model for $\alpha^e(\omega)$. The quadrupole term added in Eq.(\ref{eq:7}) allows us to calculate the near-field heat transfer between nanoparticles at shorter distances than calculations from other models \cite{5,8,25}.

The heat interchange between two particles is calculated \cite{6}
\begin{eqnarray}
Q_{ij}^{TE,TM}(\omega)=\int_0^{\infty}d\omega\left[P^{e,m}_{i\rightarrow j}(\omega)- P^{e,m}_{j\rightarrow i}(\omega) \right],
\label{eq:6}
\end{eqnarray}

The heat transfer per unit area from the top array to the bottom array is calculated
\begin{eqnarray}
Q = \sum_{i=1}^{N_1}\sum_{j=N_1+1}^{N_1+N_1}\left(Q^{TE}_{ij}+Q^{TM}_{ij}\right)/S
\label{eq:8}
\end{eqnarray}
where $N_1 = 400$ is the number of NPs in top and bottom object, $S$ is the area of an array, $Q_{TE}$ and $Q_{TM}$ are the radiative heat transfer of electric and magnetic contribution in NPs, respectively. The first and second sum correspond to the summation of nanoparticles in the bottom and top layer.

\section{Numerical results and discussions}
Increasing the distance $d$ leads to the increase of center-center distance between particles in the systems. The importance of the many-particle effect significantly reduces. Therefore, in our paper, we chose $d = 2R + 1$ nm to be suitable with pervious experiments \cite{2} and clearly exhibit the many-body effects.

\begin{figure}[htp]
\includegraphics[width=9cm]{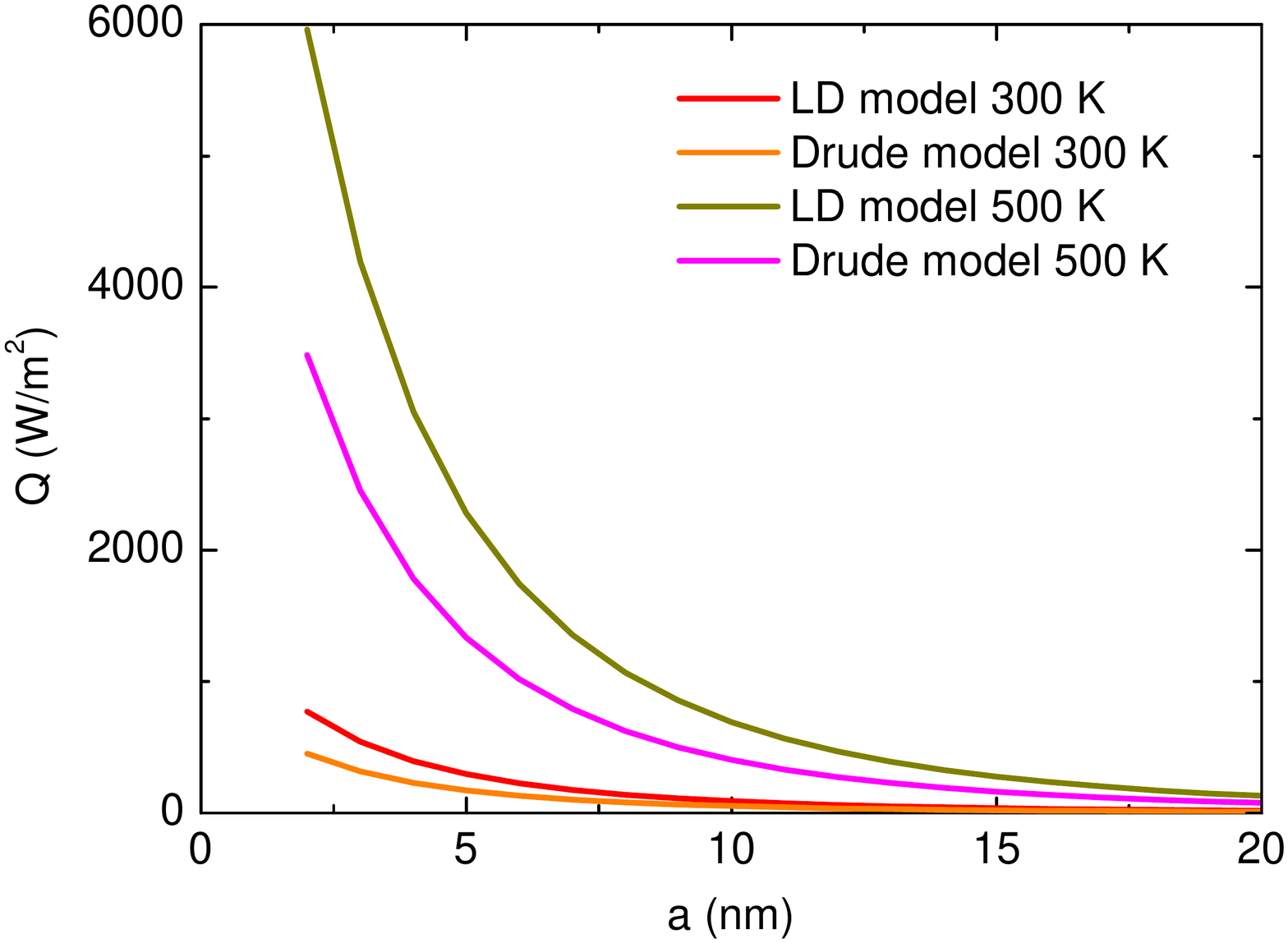}
\caption{\label{fig:1}(Color online) The radiative heat transfer between two gold MNP layers as a function of separation distance $a$ at different temperatures $T$ using the Lorentz-Drude and Drude model for the dielectric function.}
\end{figure}

The dielectric function of gold NPs is modelled by the Lorentz-Drude (LD) model \cite{11}
\begin{eqnarray}
\varepsilon(\omega)=1-\frac{f_0\omega_p^2}{\omega(\omega +i\Gamma_0)}+\sum_{j}\frac{f_j\omega_p^2}{\omega_j^2-i\omega\Gamma_j-\omega^2},
\label{eq:9}
\end{eqnarray}
where $f_0$ and and $\omega_p$ are $0.845$ and $9.01$ eV, respectively. Also, $f_j$ are the oscillator strengths corresponding to characteristic frequencies $\omega_j$ and damping parameters $\Gamma_j$ given in \cite{11}. These parameters were fitted from data set that was measured for gold nanostructure. The first two terms in Eq.(\ref{eq:9}) describe the contribution of a free electron gas to the response, while the other terms represent interband transitions. In previous studies, authors used the Drude model $\varepsilon(\omega)=1-\omega_p^2/\omega(\omega +i\Gamma_0)$ for the dielectric function of gold. The model is suitable for the dielectric response of bulk, however. The inclusion of the Lorentz oscillators accounts for the localized surface plasmon modes of MNPs with wavelengths $\sim$ 500 nm. Note that the finite spherical size of the nanoparticles affects the damping parameter $\Gamma_0$ for gold. Here we take that $\Gamma_0\rightarrow  \Gamma_0 + Av_f/R$ \cite{12}. For gold, the parameter  $A \approx 1$  and $v_f$ is the Fermi velocity of gold \cite{12}.  

We note that the finite size of the nanoparticles, taken via the modification in $\Gamma_0$, can play an important role in the heat exchange process. Fig. \ref{fig:1} shows a comparison between the heat transfer between two MNP arrays using the LD and Drude model. The bottom layer is kept at $T_0=0$ $K$, while the top layer is maintained at a finite temperature $T$. \cite{23}. For the two chosen temperatures, $Q$ is much larger for the LD model. The huge difference for two models shows that it is impossible to obtain correct value with the Drude model because of the neglect of the bound electron contribution in the polarizability.  
\begin{figure}[htp]
\includegraphics[width=9cm]{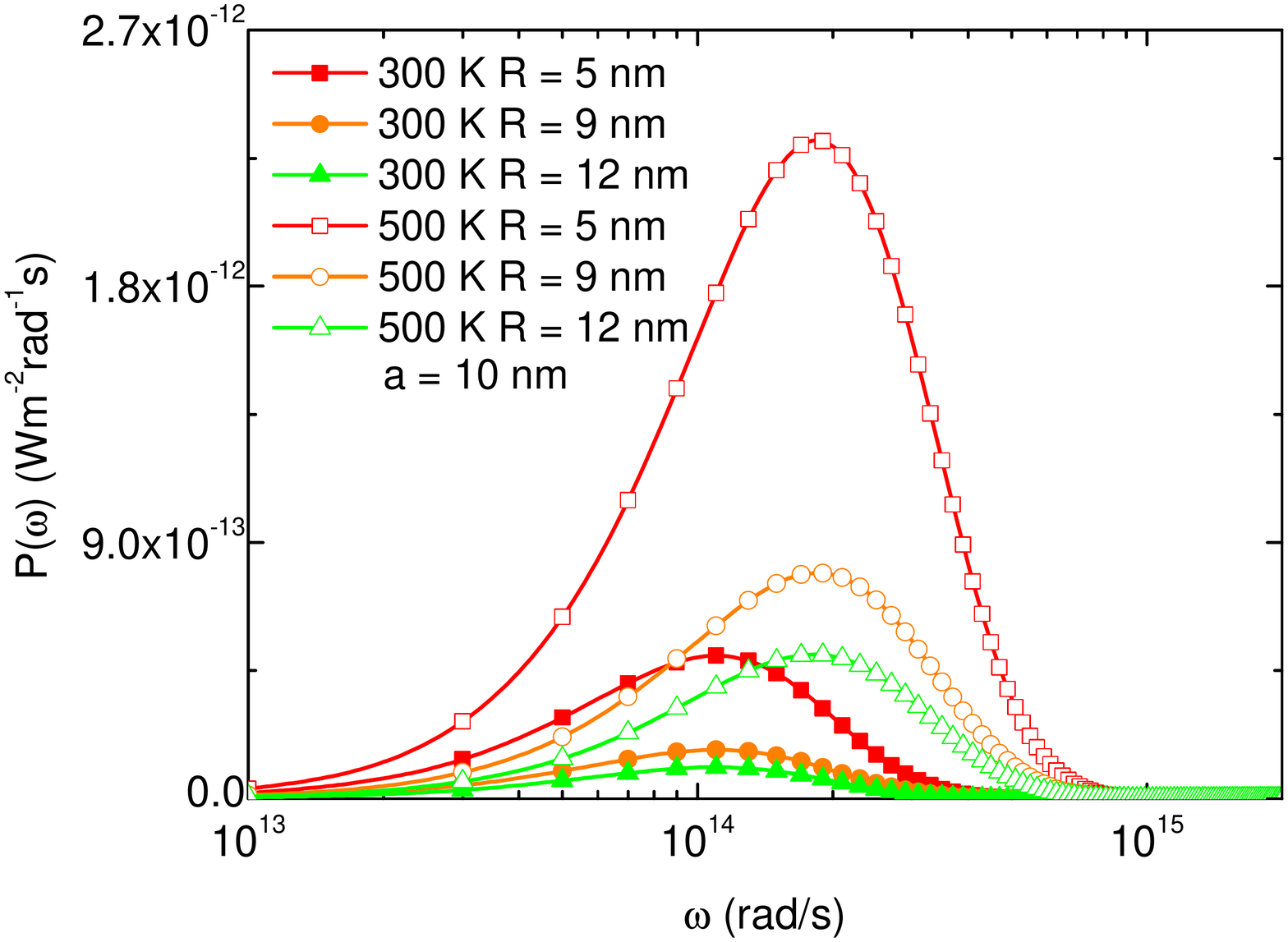}
\caption{\label{fig:5}(Color online) The heat flux between two gold nanoparticle layers as a function of $\omega$ with a variety of $R$ and $T$ at $a = 10$ nm.}
\end{figure}

To investigate the radiative heat transfer, we have to know the frequency range that is important for the thermal conductance through the heat flux as a function of frequency. The expression of the heat transfer between two arrays versus $\omega$ is given
\begin{eqnarray}
P(\omega) = \sum_{s=e,m}\sum_{i=1}^{N_1}\sum_{j=N_1+1}^{N_1+N_1}\left[ P^{s}_{i\rightarrow j}(\omega)- P^{s}_{j\rightarrow i}(\omega)\right]
\end{eqnarray}
where $N_1 = 400$ is the number of nanoparticles in a layer. The first sum corresponds to the two modes (TE, TM), the second one - to the number of particles in the top layer, and the third one - to number of particles in the bottom layer. Figure \ref{fig:5} shows the heat transfer versus frequencies with different sizes of NPs. The radiative heat transfer is contributed significantly by frequencies ranging from $2\times 10^{13}$ to $6\times 10^{14}$ rad/s. The position of the peak of $P(\omega)$ shifts from left to right when enlarging the nanoparticle's radius.

\begin{figure}[htp]
\includegraphics[width=9cm]{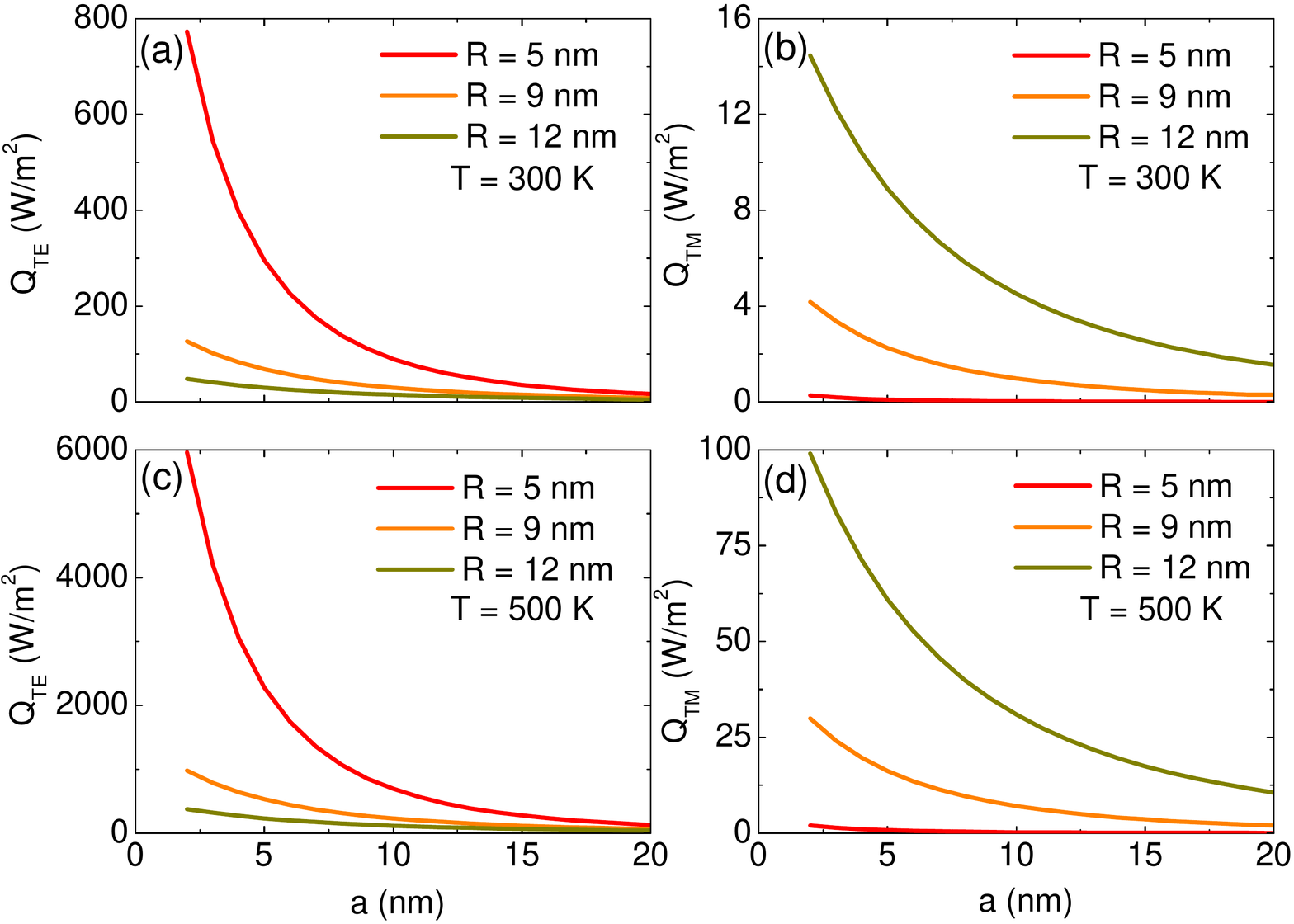}
\caption{\label{fig:2}(Color online) The heat exchange due to magnetic dipole $Q_{TM}$ and electric dipole $Q_{TE}$ contribution at temperature $T = 300$ and 500 $K$ are calculated for different MNPs with different radii.}
\end{figure}

We also investigate how $Q$ is affected by the $TE$ and $TM$ modes of the system. Fig.\ref{fig:2} shows that for spheres with smaller $R$, $Q_{TE}$ is dominant. As the radius is increased, the contribution from $Q_{TM}$ becomes more significant. The role of the quadrupole term in the electric polarizability in the absorption and scattering spectrum of MNPs becomes considerable when the NP radius is large \cite{19} because of the proportionality of the term to $R^5$. The higher-order multipole terms are found to be proportional to $R^{2l+1}/[\varepsilon + (l+1)/l]$ with the integer $l \ge 3$. Nevertheless, the quadrupole and higher-order multipole contribution to the heat transfer for the studied structures are small. This is due to the large denominator $[\varepsilon + (l+1)/l]$ and small radius $R$. The contribution of the magnetic polarizability to the heat radiation surpasses that of the quadrupole term. Using Eq. (\ref{eq:3}), (\ref{eq:5}) and (\ref{eq:7}), one finds that $Q_{TM}\sim R^{10}$ and $Q_{TE}\sim R^{6}$. Thus increasing the MNP radius enhances the effect of the magnetic polarizability and reduces the influence of the electric polarizability in the near-field radiation. At certain temperature $T$ and separation distance $a$, the heat radiation between the two nanoparticles is amplified as $R$ increases. In the layered systems, however, the heat flux $Q_{TM}$ and $Q_{TE}$ dramatically decreases because the distance from a particle to particle located in different layers, except for the nearest neighbors, increases. In comparison with bulk material and thin film systems, the near-field radiation of the nanoparticle arrays is weaker. The main reason is that the layer systems have a thin thickness and spacing between among MNPs in the same array. The total heat flux $Q$ is $1.92, 1.82$ and $1.78$ times greater than the heat flux of $400$ nearest neighbor pairs of particles placed two arrays at $a = 2$ nm for $R = 5, 9, 12$ nm, respectively. The ratios decrease when the separation $a$ is expanded since the many-body effects are strengthened if ${r}_{ij}/a$ is smaller, here $\mathbf{r}_{i}$ and $\mathbf{r}_{j}$ are the positions of particles in different layers, and $\mathbf{r}_{ij} = \mathbf{r}_{i}-\mathbf{r}_{j}$.

\begin{figure}[htp]
\includegraphics[width=9cm]{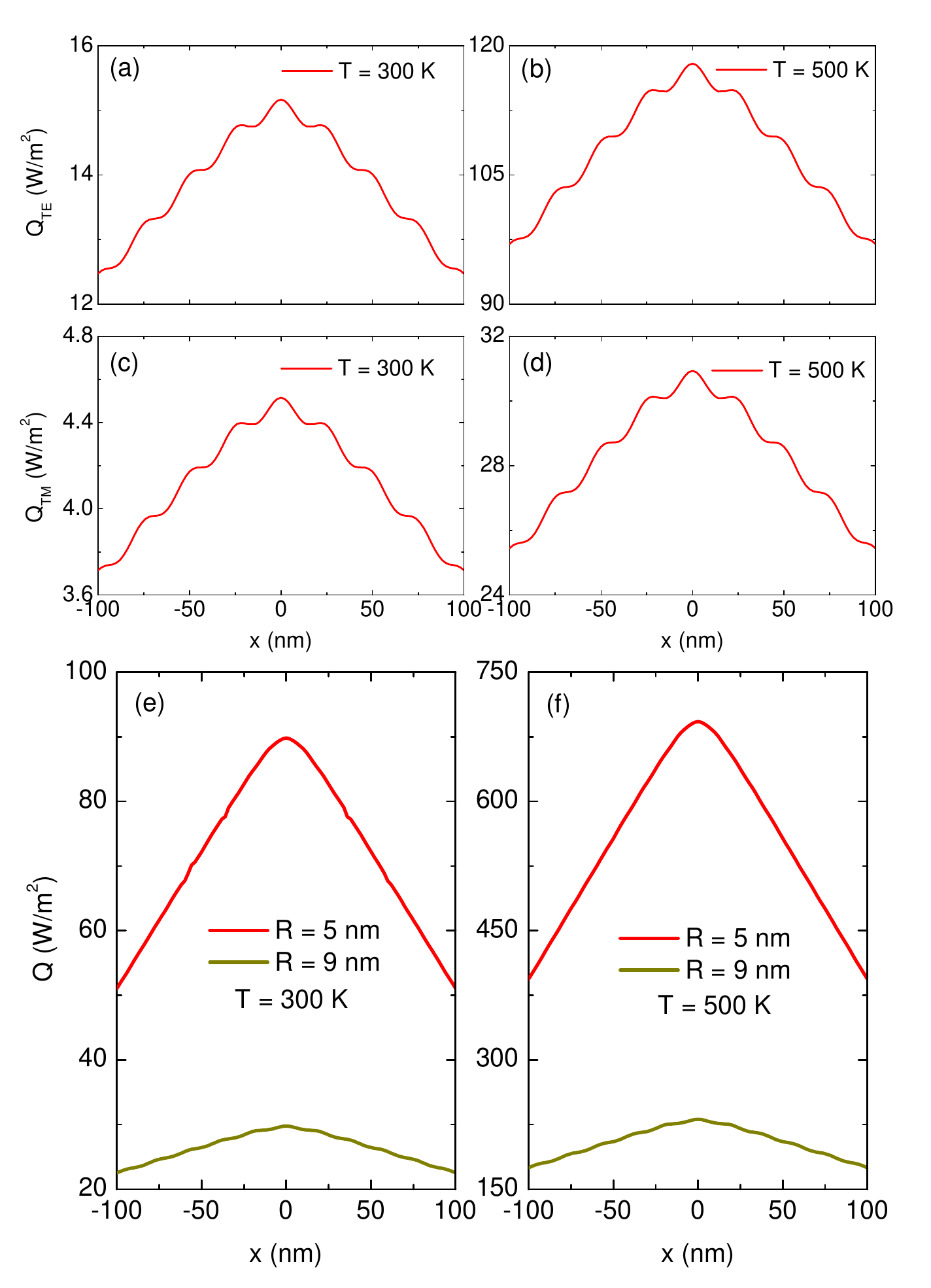}
\caption{\label{fig:3}(Color online) The radiative heat transfer $Q_{TE}$ and $Q_{TM}$ at $a = 10$ nm as a function of displacement along $x$ axis of the top gold MNP layers with $R$ = 12 nm shown in (a), (b), (c) and (d) at $300$ and $500$ $K$. The net heat flux versus $x$ with $R$ = $5$ and $9$ nm described in (e) and (f), respectively, at $300$ and $500$ $K$.}
\end{figure} 

In Fig.\ref{fig:3}, we show results for the heat transfer for the $TE$ and $TM$ modes when there is relative translational displacement along the $x$ axis between the two MNP layers.  It is found that the maximum heat is transferred when the layers are completely overlapping $(x=0)$. As the relative displacement between the layers is increased, $Q_{TE}$, and $Q_{TM}$ decrease at an oscillatory-like fashion. One finds that the period of oscillations of 25 nm for the $R=12$ nm spheres corresponds to distance separation between two neighboring nanoparticles in a layer.   

Combining the contributions from both modes, it is found that the oscillatory-like behavior of $Q$ vs $x$ is not as pronounced, although some oscillations are seen for the the nanoparticles with radius $R=9$ nm (Fig.\ref{fig:3} $e$ and $f$). Our calculations indicate that the heat transfer depends strongly on the overlap between the two layers when sliding one array along $x$ axis with respect to each other. The oscillatory trends of $Q_{TE}$ and $Q_{TM}$ for NPs $R = 12$ nm are observed by means of the couple dipole method in Fig.\ref{fig:3} (a), (b), (c) and (d). It is very easily to see that the period of this oscillatory behavior between two neighboring peaks is approximately $25$ nm, which relatively corresponds to the distance $d$ between two nearest NPs at the same array. It suggests that the oscillatory feature depends on how well the horizontal plane projections of the top gold NP array and the bottom one matches each other. For $R = 9$ nm, Fig.\ref{fig:3} (c) and (d) still show the periodic oscillation in the heat transfer band although this behavior is quite small. Thus one can conclude that when $a \gg R$, the actual distribution of the nanoparticles is not important, however, the overlap between the layers can change $Q$ by several orders of magnitude.

\section{Conclusions}
This paper has presented theoretical calculations for the near-field radiation in systems involving gold MNPs. Our method can investigate the discrete nanostructures with arbitrary geometries and consider the size effect of NPs including in the dielectric response. We have considered the role of the structure of MNP layers on the heat transfer when these two arrays are displaced with respect to each other along parallel and perpendicular directions. These results can provide guidelines for designing thermal devices utilizing electromagnetic radiation.
\begin{acknowledgments}
Lilia M. Woods acknowledges the Department of Energy under contract DE-FG02-06ER46297.
\end{acknowledgments}

\end{document}